\documentclass[aps,prl,preprint]{revtex4-1}
\pdfoutput=1

\usepackage[a4paper,centering]{geometry}
\usepackage[pdftex]{graphicx} 
\usepackage{hyperref}

\usepackage[format=plain,justification=raggedright,singlelinecheck=false,font=small,labelfont=bf,labelsep=space]{caption}

\begin{document}

\noindent\LARGE{\textbf{High-order harmonic spectroscopy\\for molecular imaging 
of polyatomic molecules}}
\vspace{0.6cm}

\noindent\large{\textbf{M. Negro,\textit{$^{a}$} M. Devetta,\textit{$^a$} D. 
Faccial\'{a},\textit{$^b$} S. De Silvestri,\textit{$^b$} C. Vozzi\textit{$^{a 
\ast}$} and S. Stagira,\textit{$^b$}}}\vspace{0.5cm}

\vspace{0.5cm}
\noindent\footnotesize{
  \\{\textit{$^{a}$~Istituto di Fotonica e Nanotecnologie - CNR, 20133 Milan, Italy}}
  \\{\textit{$^{b}$~Dipartimento di Fisica - Politecnico di Milano, 20133 Milan, Italy}}
  \\{$^{\ast}$ corresponding author: caterina.vozzi@ifn.cnr.it}
}

\vspace{0.6cm}

\noindent \normalsize High-order harmonic generation is a
powerful and sensitive tool for probing atomic and molecular structures,
combining in the same measurement an unprecedented attosecond temporal
resolution with a high spatial resolution, of the order of the angstrom.
Imaging of the outermost molecular orbital by high-order harmonic generation has 
been limited for a long time to very simple molecules, like nitrogen.
Recently we demonstrated a technique that overcame several of the issues that 
have prevented the extension of molecular orbital tomography to more complex
species, showing that molecular imaging can be applied to a triatomic molecule
like carbon dioxide. Here we report on the application of such technique to
nitrous oxide (N$_2$O) and acetylene (C$_2$H$_2$). This result represents a 
first step towards the imaging of fragile compounds, a category
which includes most of the fundamental biological
molecules.

\vspace{0.5cm}

\section{Introduction\label{intro}}

High order harmonic generation (HHG) occurs when atoms or 
molecules exposed to an intense femtosecond laser pulse are ionized by
tunneling. The freed electron is then accelerated in the external electric
field. Because of the periodic oscillation of the laser field, the electron is
brought back to the parent ion where it may recombine emitting an XUV photon
\cite{corkum1993}. This XUV radiation has been shown to contain information on
the electronic structure of the emitting molecule and on its internal dynamics.
Attosecond nuclear \cite{baker2006} and electronic dynamics \cite{smirnova2009, 
haessler2010} have been extracted from HHG in simple molecules and spectral 
features in the harmonic emission have been related to the molecular electronic
structure and have been used for imaging the highest occupied molecular orbital
(HOMO).

The idea of exploiting HHG for the tomographic reconstruction of 
molecular orbitals was first introduced by Itatani et al. in 2004 for the 
nitrogen molecule \cite{itatani2004}.
Since then, numerous experiments have been realized, addressing the role of the 
HOMO in the harmonic spectral intensity \cite{kanai2005,vozzi2005},
in the molecular-frame photo-ionization \cite{le2009a} and in the subsequent 
attosecond XUV emission \cite{boutu2008}, as well as in the polarization state
of the emitted radiation \cite{levesque2007}.
The dependence of the HHG process on the HOMO structure has also been exploited 
for the characterization in the time domain of the rotational
\cite{miyazaki2005} and vibrational \cite{li2008} molecular excitations.

All these studies rely on two major assumptions: (i) the 
molecular HHG is dominated by the HOMO structure; (ii) the relationship between
molecular structure and emitted XUV spectrum is simple and completely captured
by the Strong Field Approximation (SFA), i.e. the electron quiver motion is not
perturbed by the Coulomb potential of the ion.

Both these assumptions have been recently put into question.
Recent experiments enlightened the role of multiple orbital contributions to HHG
emission \cite{smirnova2009,haessler2010}. Furthermore, the influence of the
Coulomb field of the parent ion in the generation of high order harmonics from
molecules has been considered as a serious hindrance to a clear HOMO
reconstruction \cite{walters2008}. These assumptions should then be overtaken to
perform molecular tomography to more complex species.

Besides these two more fundamental obstacles, there are also
additional, more technical difficulties. In order to retrieve the HOMO
structure, one has to record the XUV harmonic spectra for different molecular
orientations with respect to the laser field. Hence, it is necessary to fix the
molecular orientation in space and change the polarization direction of the
HHG-driving field \cite{itatani2004}. 
Laser-assisted molecular alignment is a widespread technique able to accomplish 
this task \cite{stapelfeldt2003}, but the molecular alignment achieved in this 
way is not ideal. Hence the experimental results and the corresponding HOMO 
tomography are affected by angular averaging effects. Moreover, in the case of 
non-linear molecules, the tomographic procedure requires to fix two or three 
angular coordinates of the molecule under investigation. For instance, the study 
of linear polar molecules requires to fix the head-tail direction in space. The 
feasibility of laser assisted molecular orientation has been recently 
demonstrated \cite{le2009b} and exploited in HHG spectroscopy
\cite{frumker2012a,frumker2012b,spanner2012}, but no direct application to
molecular imaging has been yet realized.

The amount of information that can be extracted from the
harmonic emission depends on the spectral extension of the XUV radiation, that
is known to scale with the so-called cut-off law: $E_{max} = I_p + 3.17U_p$,
where $I_p$ is the ionization potential of the molecule and $U_p$ is the
ponderomotive energy of the electron in the laser field. This poses another
important problem when HHG molecular imaging is extended to species with low
ionization potential (i.e. all organic molecules, and in particular those having
important biological functions) as the extension emission spectrum is reduced.
Since $\mathrm{U_p} \propto \lambda^2 \mathrm{I}$, where I is the peak intensity 
and $\lambda$ the wavelength of the driving laser pulse, the emission cut-off
may be extended by both increasing the field intensity or the laser wavelength.
In this respect, standard Ti:Sapphire laser sources generally used in HHG are
not ideal candidates for tomography in fragile molecules, since the intense
optical fields needed completely ionize the molecule before a well-developed
XUV spectrum is generated.

To overcome the limitations posed by ionization saturation,
the exploitation of mid-infrared driving sources has been demonstrated to be a
powerful tool to extend harmonic emission far in the XUV
\cite{takahashi2008,vozzi2010,vozzi2010b,vozzi2010c,popmintchev2012}.

With a mid-IR source\cite{vozzi2007} we recently demonstrated
that it is possible to extend the spectral investigation in carbon dioxide
beyond 100 eV in the absence of multielectron effects, thus avoiding any
ambiguity in the reconstructed wavefunction. In addition, by exploiting an
all-optical non-interferometric technique, it was possible to trace both the
spectral intensity and phase of high order harmonics generated by single
molecules as a function of emitted photon energy and molecular angular
orientation, without averaging effects. Furthermore, the tomographic procedure
was generalized in order to take into account the Coulomb potential seen by the
re{-}colliding electron wavepacket\cite{vozzi2011}.

In this work, we extend that approach to more complex molecules, such as N$_2$O
and C$_2$H$_2$ pointing out some strengths and weaknesses of this investigation
technique.

\section{Experimental Setup\label{setup}}
We exploited an optical parametric amplifier (OPA) pumped by an amplified
Ti:sapphire laser system (60 fs, 20 mJ, 800 nm). The OPA is based on difference
frequency generation and provides driving pulses with 1450 nm central
wavelength, pulse duration of 20 fs and pulse energy of 1.2 mJ\cite{vozzi2007}.
High harmonics were generated by focusing the mid-IR pulse in a
supersonic gas jet under vacuum, due to the strong absorption exhibited by air
in the XUV spectral region.
The molecules in the jet were impulsively aligned with a portion of the
fundamental 800-nm beam which was spectrally broadened by optical filamentation
in an argon-filled gas cell and temporally stretched up to 100 fs by propagation
through a glass plate. Such duration is required for achieving a good alignment
of the molecular sample.
In our experimental setup, driving and aligning pulse were collinear and their
polarizations were parallel. The delay between the two pulses was adjusted by
means of a fine-resolution translation stage. The XUV radiation was acquired
by means of a flat-field spectrometer and a multi-channel plate detector coupled
to a CCD camera \cite{poletto2001}.

\section{Results\label{results}}

\begin{figure}[!ht]
\centering
\includegraphics[width=0.8\textwidth]{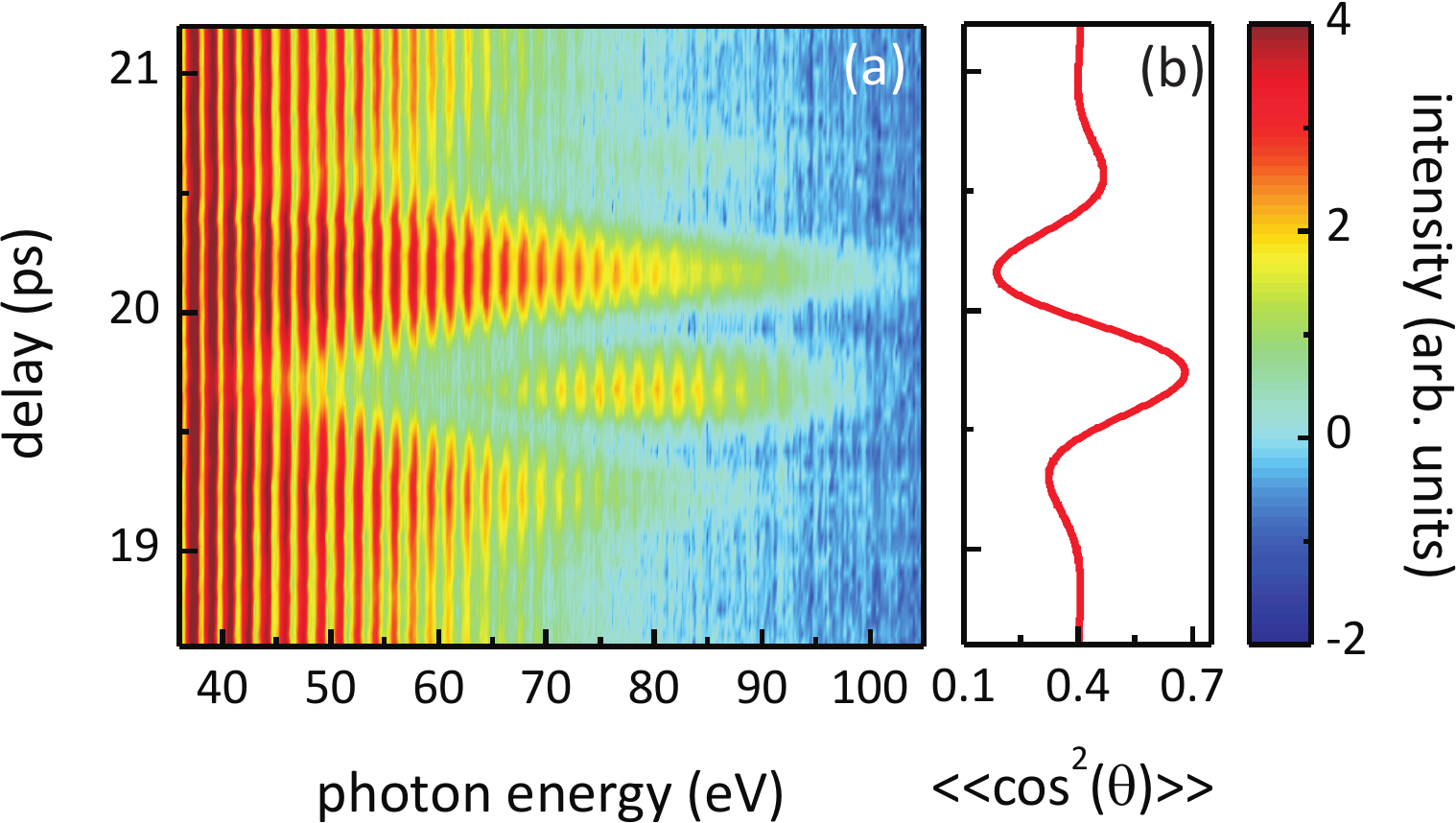}
\caption{(a) Sequence of harmonic spectra measured in N$_{2}$O as a function of 
emitted photon energy and delay between the aligning and the driving pulse (log
scale). (b) Calculated alignment factor for N$_{2}$O in the experimental
conditions (rotational temperature 75 K, aligning pulse duration 100 fs,
aligning pulse intensity $3.32\times 10^{13}$ W/cm$^2$). \label{hhg_N2O}}
\end{figure}

\begin{figure}[!ht]
\centering
\includegraphics[width=0.8\textwidth]{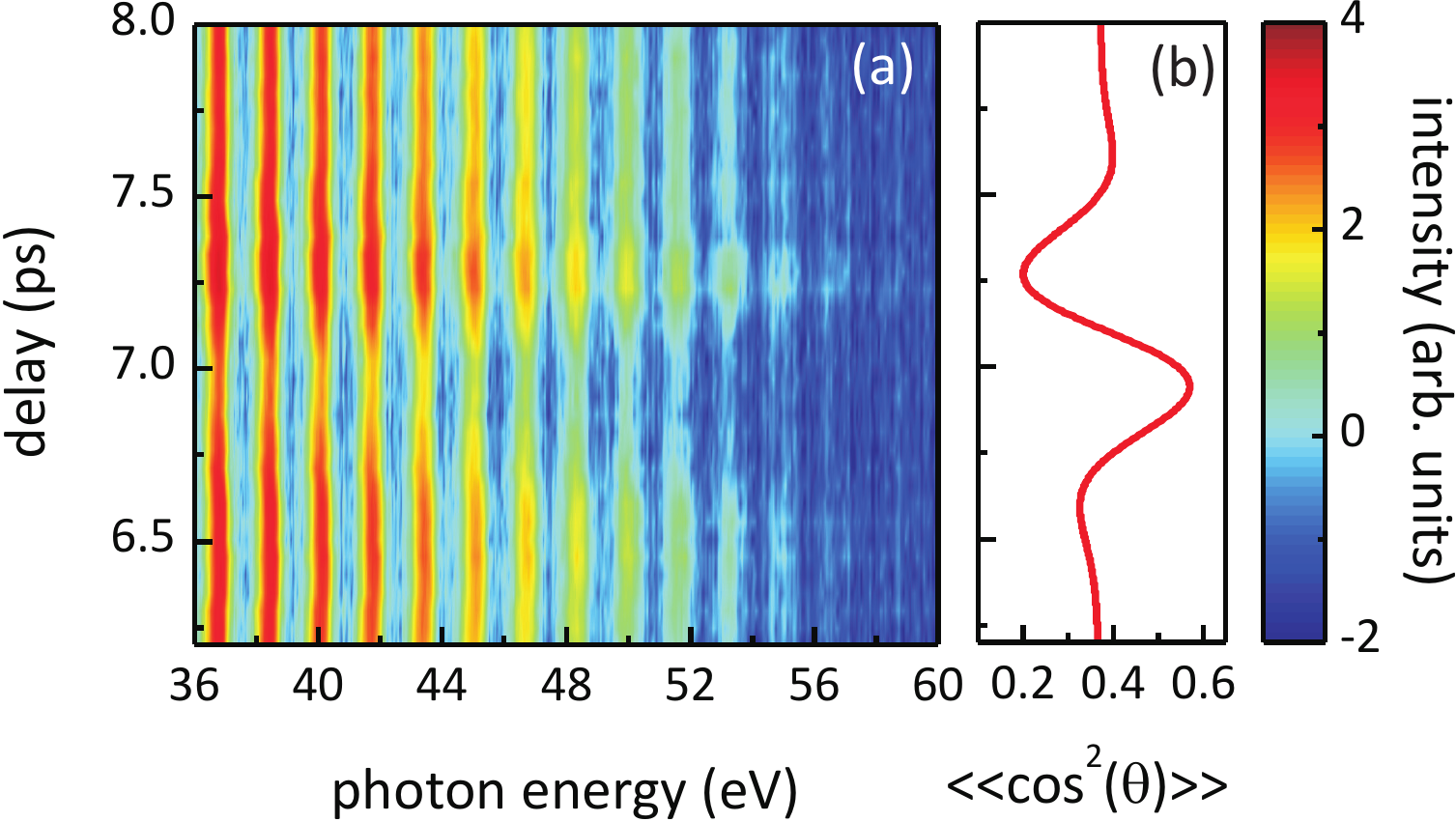}
\caption{(a) Sequence of harmonic spectra measured in C$_{2}$H$_{2}$ as a 
function of emitted photon energy and
delay between the aligning and the driving pulse (log scale). (b) Calculated 
alignment factor for C$_{2}$H$_{2}$ in the experimental conditions (rotational 
temperature 75 K, aligning pulse duration 100 fs, aligning pulse intensity 
$2.16\times 10^{13}$ W/cm$^2$). \label{hhg_C2H2}}
\end{figure}

Harmonic spectra were acquired in N$_{2}$O and C$_{2}$H$_{2}$ as a function of 
the delay $\tau$ between the aligning and driving pulse around the first 
rotational half revival ($\tau_{N_2O}=19.95$ ps and $\tau_{C_2H_2}=7.08$ ps). 
The results are shown in figure \ref{hhg_N2O}(a) and \ref{hhg_C2H2}(a) for 
N$_{2}$O and C$_{2}$H$_{2}$ respectively. Figures \ref{hhg_N2O}(b) and 
\ref{hhg_C2H2}(b) show the corresponding calculated alignment factor for the 
experimental conditions.

In both molecules, the sequence of harmonic spectra shows a strong modulation 
with the delay $\tau$ that can be ascribed to the dependence of harmonic yield 
on the molecular orbital structure. In particular, a reduction of the harmonic 
emission can be observed for the delay corresponding to the maximum of the 
alignment factor and an enhancement of the harmonic yield appears for the 
minimum of the alignment factor. A major difference between the two cases is 
the presence of a region of harmonic enhancement at high photon energy, that 
appears in N$_2$O at maximum alignment.

These effects can be naively interpreted in terms of two-center interference 
occurring in the re-collision step \cite{lein2002, vozzi2005}. If one consider a 
diatomic homo-nuclear molecule with a symmetric electronic state with respect 
to the nuclei exchange and assumes the re-colliding electron as a plane wave,
the condition for constructive interference reads $R\cos(\theta) = n \lambda_B$, 
where $R$ is the internuclear separation, $\theta$ is the angle between the 
molecular axis and the electron wave-vector, $n$ is an integer number and 
$\lambda_B$ is the de Broglie wavelength associated to the re-colliding electron 
wave-packet. Similarly the condition for destructive interference is 
$R\cos(\theta) = (n+1/2) \lambda_B$ and the first destructive interference 
occurs for $n=0$. The conditions become reversed for molecules with 
antisymmetric electronic structure.

This concept can be extended to the molecules subject of our investigation. The 
acetylene molecule has a symmetric $\pi$ HOMO in which the separation between 
the carbon atoms is $R_{\mathrm{C \equiv C}}=1.2\ \textrm{\AA}$. This is the 
distance that 
should be considered for the evaluation of the interference condition. The 
N$_2$O HOMO does not have a clear symmetry, however in our experimental 
condition the harmonic spectra are acquired in aligned molecules and correspond 
to the average between the two possible orientation. The resulting signal can 
be 
interpreted in terms of emission from an effective molecular orbital similar to 
the anti-symmetric $\pi$ orbital of CO$_2$. In this view the overall length of 
this ``effective'' orbital is $R_{\mathrm{N_2O}}=2.3\ \textrm{\AA}$. Since 
$R_{\mathrm{N_2O}} 
\approx 2 R_{\mathrm{C \equiv C}}$, a destructive interference occurs in the 
same spectral 
region for both molecules, corresponding to $n=1$ for N$_2$O and $n=0$ for 
C$_2$H$_2$.

Figures \ref{hhg_N2O}(a) and \ref{hhg_C2H2}(a) show two peculiar advantages 
related to the exploitation of mid-IR driving pulses for HHG. Indeed the 
harmonic cutoff extension related to the increase in the ponderomotive energy 
with respect to standard Ti:sapphire sources allows the observation of spectral 
features as the harmonic enhancement for high photon energy visible in N$_2$O 
in correspondence of the revival peak. In the framework of the above mentioned 
two-center model, this feature can be attributed to the appearance of 
constructive interference in that spectral region. Moreover, for the same 
emitted photon energy, mid-IR driving wavelengths require a lower pulse peak 
intensity thus reducing the ionization saturation in species with relatively 
low ionization potential, such as $C_2H_2$ ($\mathrm{I_P}=11.4$ eV).

\section{Reconstruction of Single Molecule XUV Emission \label{SMXEM}}

\begin{figure}[!ht]
\centering
\includegraphics[width=0.8\textwidth]{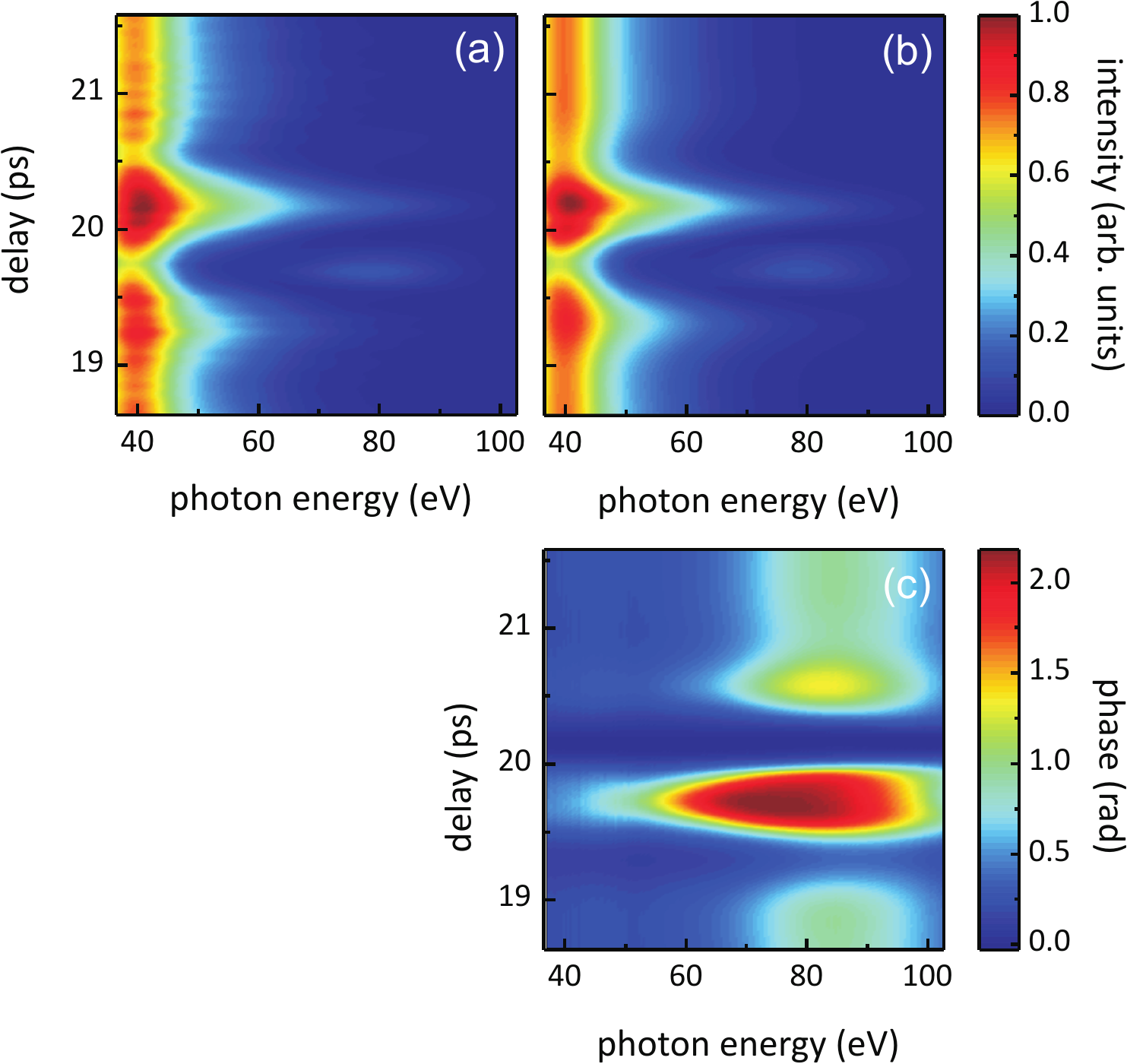}
\caption{(a) Sequence of XUV spectra measured in N$_{2}$O as a function of 
emitted photon energy and delay between the aligning and the driving pulse; the 
harmonic structure has been filtered out. Retrieved macroscopic harmonic 
emission amplitude (b) and phase (c) corresponding to the data reported in 
(a).\label{macro_N2O}}
\end{figure}

\begin{figure}[!ht]
\centering
\includegraphics[width=0.8\textwidth]{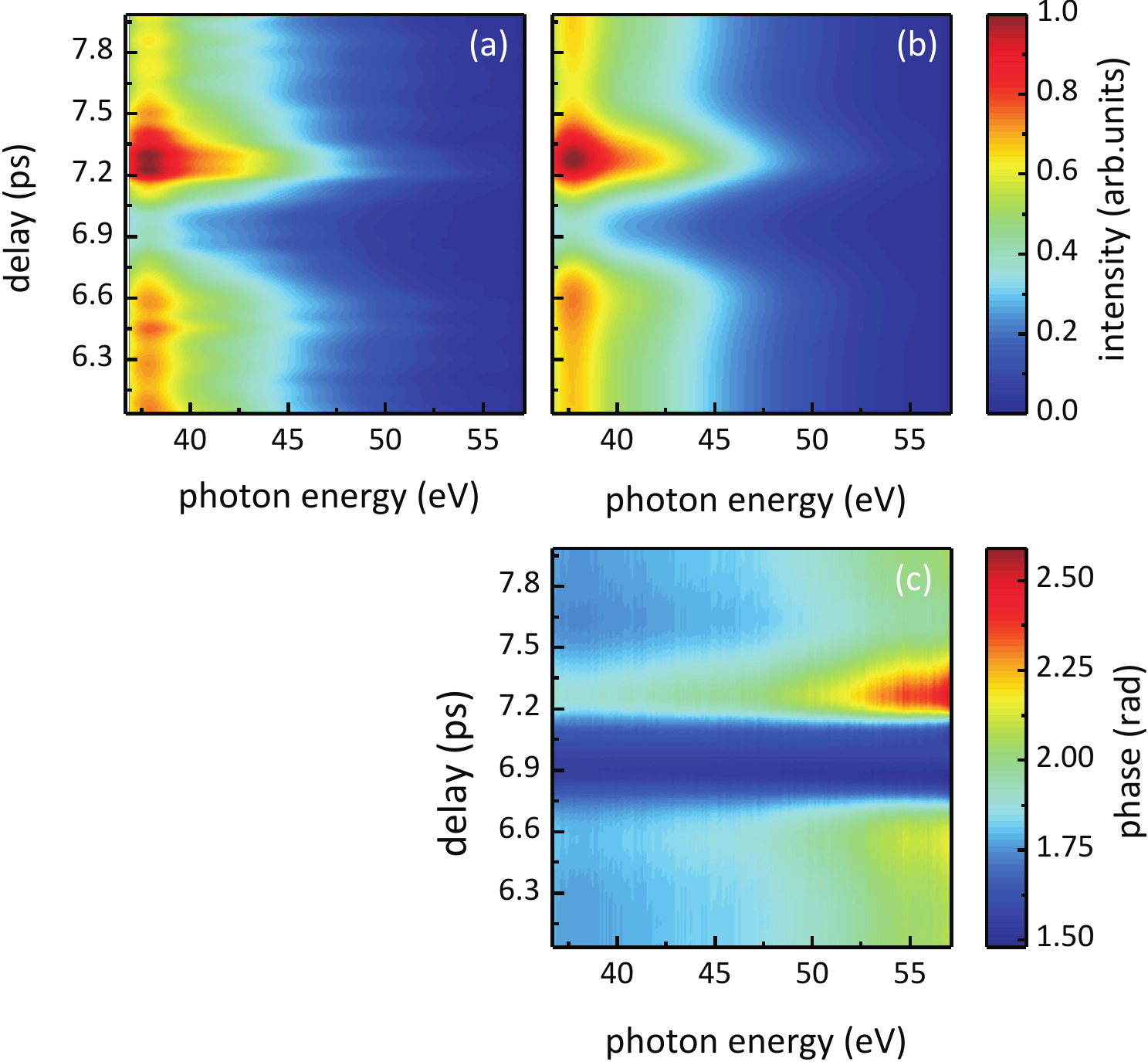}
\caption{(a) Sequence of XUV spectra measured in C$_{2}$H$_{2}$ as a function 
of 
emitted photon energy and delay between the aligning and the driving pulse; the 
harmonic structure has been filtered out. Retrieved macroscopic harmonic 
emission amplitude (b) and phase (c) corresponding to the data reported in 
(a).\label{macro_C2H2}}
\end{figure}

From the experimental data reported in figures \ref{hhg_N2O}(a) and 
\ref{hhg_C2H2}(a) it is possible to retrieve structural information on the 
target molecule following the approach introduced by Vozzi et al. 
\cite{vozzi2011}. Figures 
\ref{macro_N2O}(a) and \ref{macro_C2H2}(a) show the same experimental results 
presented in figures \ref{hhg_N2O}(a) and \ref{hhg_C2H2}(a), in which the 
harmonic structure due to the periodic re-collision of the electron wave-packet 
has been filtered out. These results have been exploited for the reconstruction 
of the XUV field emitted from a single molecule and projected on the 
polarization direction
of the aligning field  as a function of the angle between the molecular 
axis and the driving polarization direction. The reconstruction is based on a 
combination of a phase-retrieval algorithm and a Kaczmarz algorithm 
\cite{popa2004}. The main idea behind this approach is that the macroscopic XUV 
emission is the coherent superposition of the XUV field emitted by all 
molecules 
weighted with their angular distribution. This distribution changes along the 
revival in a predictable way, 
hence the sequence of harmonic emission contains enough 
information for the reconstruction of the harmonic electric field in amplitude 
and phase.

\begin{figure}[!ht]
\centering
\includegraphics[width=0.8\textwidth]{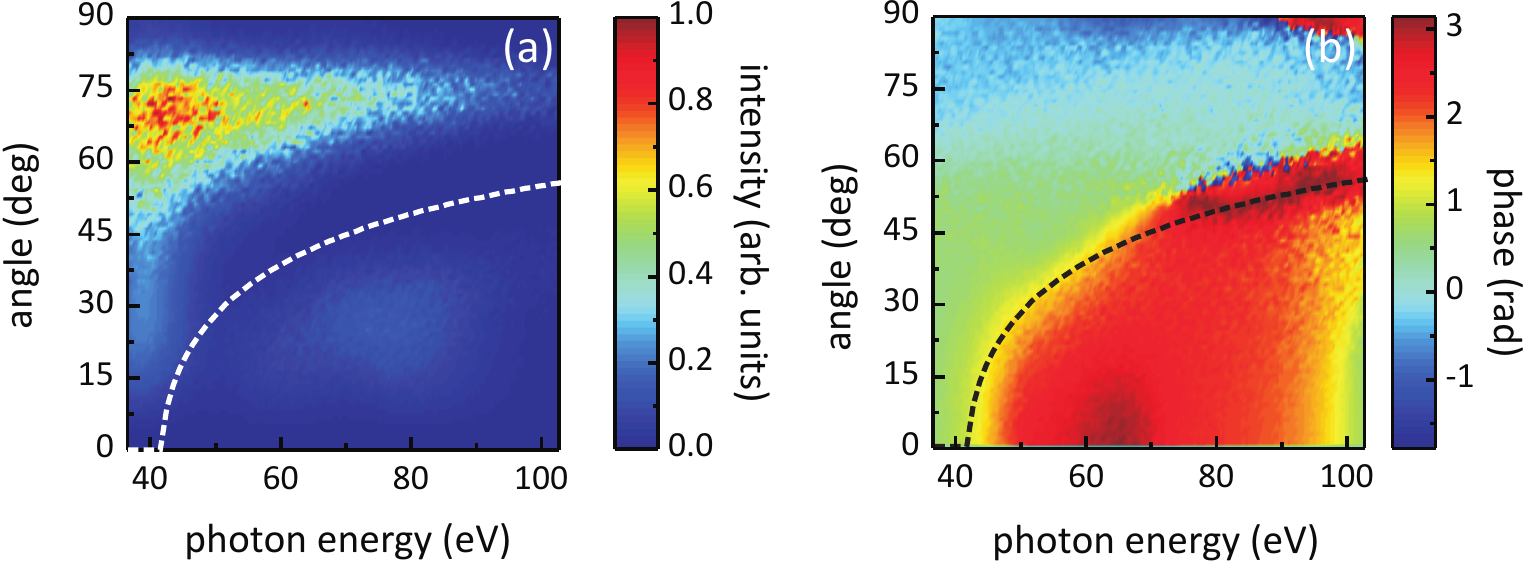}
\caption{Retrieved single molecule XUV emission map in N$_{2}$O as a function 
of 
emitted photon energy and the angle between the molecular axis and the aligning 
beam polarization direction in amplitude (a) and phase(b). Dashed lines show 
the 
position of the destructive interference predicted by the two-center model. 
\label{SMXEM_N2O}}
\end{figure}

\begin{figure}[!ht]
\centering
\includegraphics[width=0.8\textwidth]{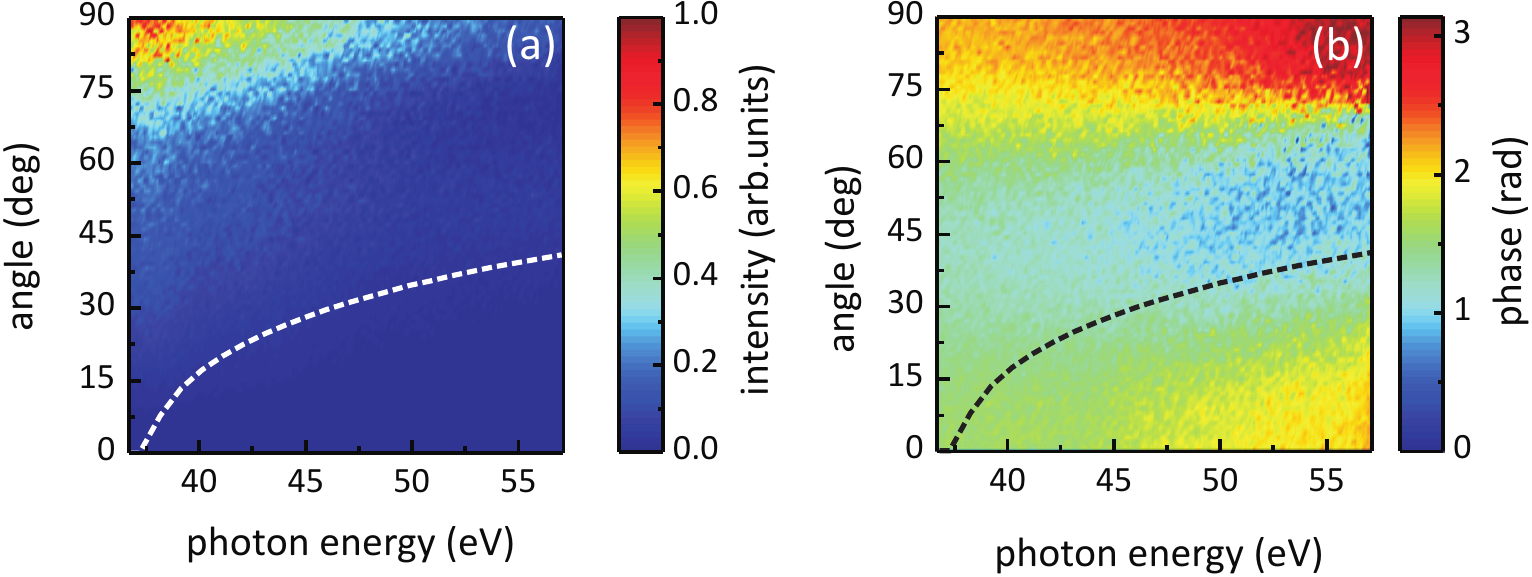}
\caption{Retrieved single molecule XUV emission map in C$_{2}$H$_{2}$ as a 
function of emitted photon energy and the angle between the molecular axis and 
the aligning beam polarization direction in amplitude (a) and phase(b). Dashed 
lines show the position of the destructive interference predicted by the 
two-center model. \label{SMXEM_C2H2}}
\end{figure}

The result of this reconstruction is shown in figure \ref{SMXEM_N2O} for 
N$_{2}$O and in figure \ref{SMXEM_C2H2} for C$_{2}$H$_{2}$. In both figures, 
panel (a) reports the amplitude of the XUV field and panel (b) shows the 
corresponding phase. In N$_{2}$O there is a clear phase jump of about 2 rad, 
that changes its position with photon energy and molecular alignment. This 
phase jump corresponds to a minimum in the XUV amplitude and its position is 
quite in good agreement with the prediction of the naive two-center model 
introduced above, which is shown as a dashed line in the figure.
It is worth noting that the reconstruction technique is based on the 
interference of XUV emission from different molecular orientations, thus the 
phase can be retrieved as a function of $\theta$ at fixed XUV photon energy. In 
order to retrieve the phase relationship between contributions at neighboring 
energies, it is necessary to introduce an a priori condition that can be 
derived 
from theoretical considerations or experimental measurements. In the case of 
N$_{2}$O we imposed a flat spectral phase of the macroscopic harmonic emission 
for the delay corresponding to the molecular anti-alignment. This condition was 
chosen in analogy with the CO$_{2}$ case\cite{vozzi2011}, due to the similarity
between the two HOMOs as discussed in the previous section.
The results of this assumption can be observed in figure \ref{macro_N2O}, where 
the reconstructed amplitude (b) and phase (c) of the macroscopic XUV emission 
from N$_{2}$O are reported. The retrieved 
amplitude is in good agreement with the experimental data (figure 
\ref{macro_N2O}(a)). The phase of the macroscopic emission shows a steep change 
of about 2 rad around 50 eV at the delay $\tau$ corresponding to the maximum 
alignment.

In the case of C$_{2}$H$_{2}$ we followed the same approach in the retrieval 
procedure.
We imposed in this case a flat spectral phase for the macroscopic harmonic 
emission
at the delay $\tau$ corresponding to the molecular alignment. 
This assumption was necessary in order to complete the retrieval procedure, but 
it is arbitrary
and not supported by theoretical models; it could be however improved by 
changing the retrieving
condition according to an experimental spectral phase measurement. 
This kind of experiment can be  performed for example by RABBIT technique at a 
given alignment delay \cite{boutu2008}. 
The retrieved single molecule XUV emission in C$_{2}$H$_{2}$, shown in figure 
\ref{SMXEM_C2H2}, is very different from the one reported for N$_{2}$O. In 
particular a strong contribution comes from molecules with perpendicular 
orientation with respect to the driving field polarization direction. In the 
retrieved phase (figure \ref{SMXEM_C2H2}(c)) two phase jumps are clearly 
observed. The first one appears for small alignment angles and roughly follows 
the prediction of the two-center model. The second jump appears at large 
alignment angles and may be attributed to the shape of the HOMO seen by the 
re-colliding electron. However, since the reconstruction is based on the 
arbitrary assumption of flat macroscopic spectral phase at the alignment delay, 
the retrieved outcomes should be considered preliminary. 
In spite of this, the retrieved macroscopic XUV amplitude (figure 
\ref{macro_C2H2}(b)) 
is in fair agreement with the experimental results.

\section{Molecular Orbital Tomography \label{tomo}}

\begin{figure}[!ht]
\centering
\includegraphics[width=0.5\textwidth]{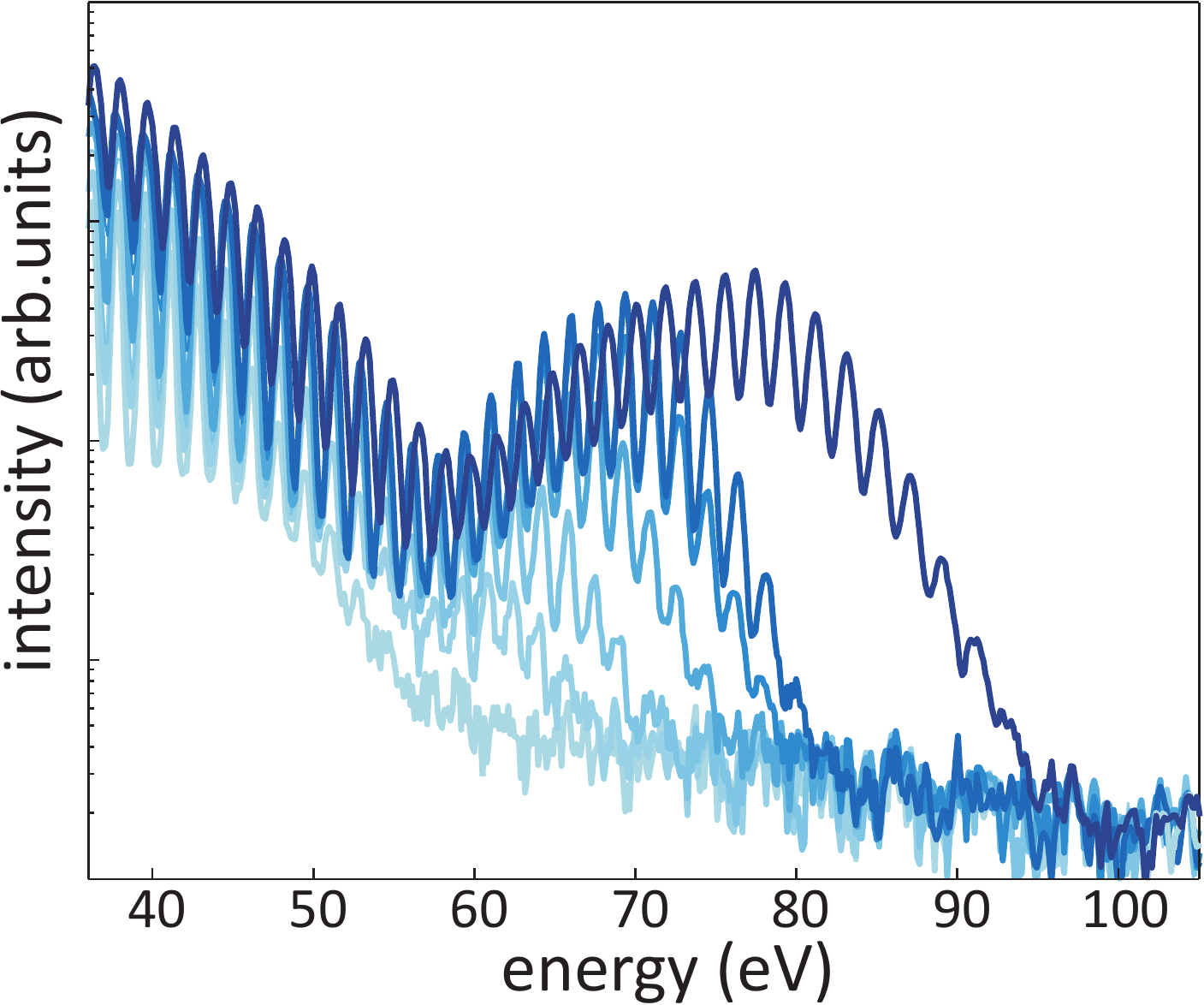}
\caption{Harmonic spectra generated in N$_{2}$O at the delay $\tau$ 
corresponding to the maximum molecular alignment for several driving peak 
intensities I between 1 and $1.7\times 10^{14}$ W/cm$^2$. \label{intscan}}
\end{figure}

\begin{figure}[!ht]
\centering
\includegraphics[width=0.8\textwidth]{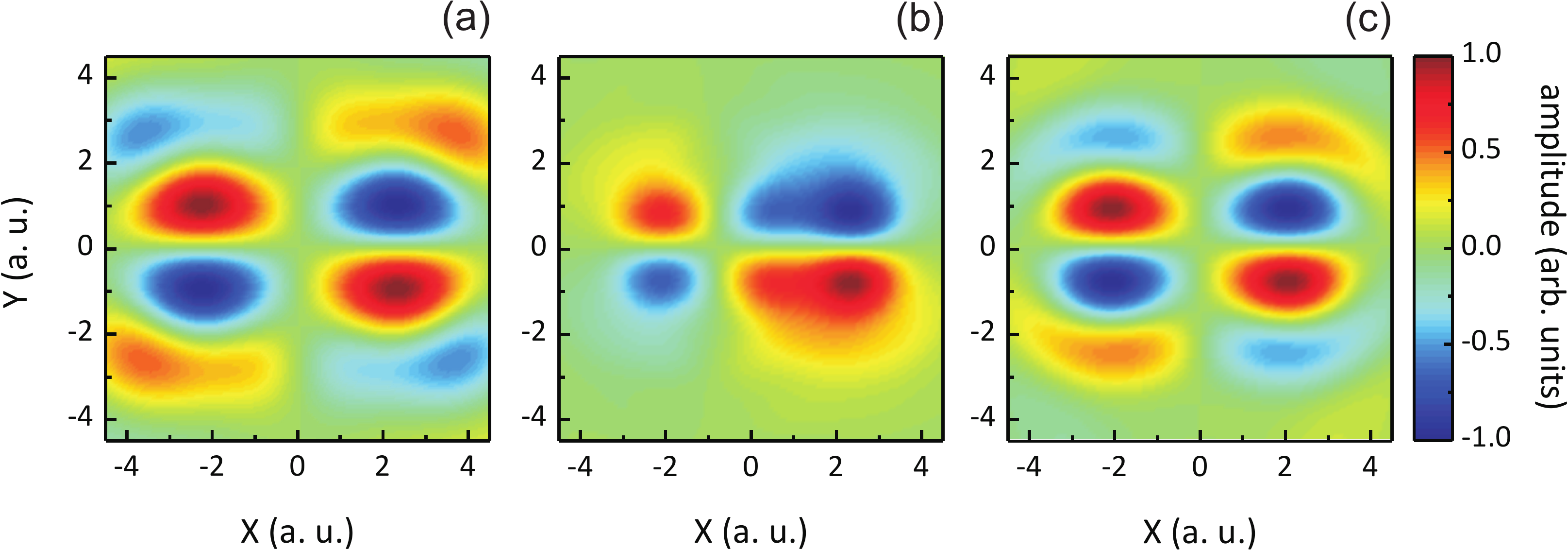}
\caption{(a) Highest occupied molecular orbital of N$_{2}$O as retrieved from 
the single molecule XUV emission map. (b) Highest occupied molecular orbital of 
N$_{2}$O calculated with a quantum chemistry program \cite{dalton}. (c) 
N$_{2}$O 
HOMO calculated averaging over the two possible orientations of the molecular 
axis and considering the filtering in spectral domain corresponding to the 
experimental conditions.\label{homo_N2O}}
\end{figure}

The results reported in the previous section can be used for the 
two-dimensional 
reconstruction of molecular orbitals, following the tomographic procedure 
proposed by Itatani et al. \cite{itatani2004} and extended by Vozzi et al. 
\cite{vozzi2011}. However to proceed with this tomographic reconstruction, it 
is necessary to rule out the occurrence of multi-electron effects in HHG. 
A simple experimental procedure to check whether spectral modulations in 
harmonic emission are due to multi-electron effects is to change the driving 
field intensity. As shown by Smirnova et al. \cite{smirnova2009}, one expects 
all the features 
due to multi-electron effects to shift with the driving field intensity. Figure 
\ref{intscan} shows harmonic spectra acquired in aligned N$_{2}$O for a delay 
$\tau$ corresponding to the maximum of the alignment for different values of 
the driving intensity. The spectral minimum associated to the phase change 
retrieved in figure \ref{SMXEM_N2O}(b) appears always around 55 eV and does not
shift with the intensity. This behavior guarantees that the main spectral features in the 
harmonic emission are mainly dictated by the HOMO structure. This consistency 
check allowed us to exploit the retrieved single molecule harmonic emission for 
the reconstruction of N$_{2}$O orbital. The result is shown in figure 
\ref{homo_N2O}(a). Figure \ref{homo_N2O}(b) shows the  N$_{2}$O orbital 
calculated with a quantum chemistry program \cite{dalton}. Even if the overall 
dimension of the 
molecular orbital is well reproduced, the asymmetry of this orbital is very 
clear and cannot be addressed by the tomographic reconstruction, since in the 
experiment the molecules were aligned but not oriented. Another departure of 
the retrieved orbital with respect to the calculated one is the presence of side 
lobes, that can be attributed to the limited working range of the XUV 
spectrometer used in these experiments. Since there is a correspondence between 
the energy range of harmonic emission and the spatial frequency domain, the 
limited spectral range collectible in the experiment corresponds to a spatial 
filtering in the Fourier domain, which gives raise to such lobes. These 
observations are further confirmed by figure \ref{homo_N2O}(c), which shows the 
calculated HOMO corresponding to the average between the two possible 
orientations of N$_{2}$O molecular axis and takes into account the limited 
spectral bandwidth available in 
the experiment. The features of this fictitious orbital are in very good 
agreement with the reconstruction of figure \ref{homo_N2O}(a).
It is worth nothing that such limitations can be overcome by extending the 
acquired spectral range over all the XUV emission and by exploiting all-optical 
impulsive techniques for orientation of polar molecules, such the one 
demonstrated by Frumker et al.\cite{frumker2012a,frumker2012b}.

\begin{figure}[!ht]
\centering
\includegraphics[width=0.8\textwidth]{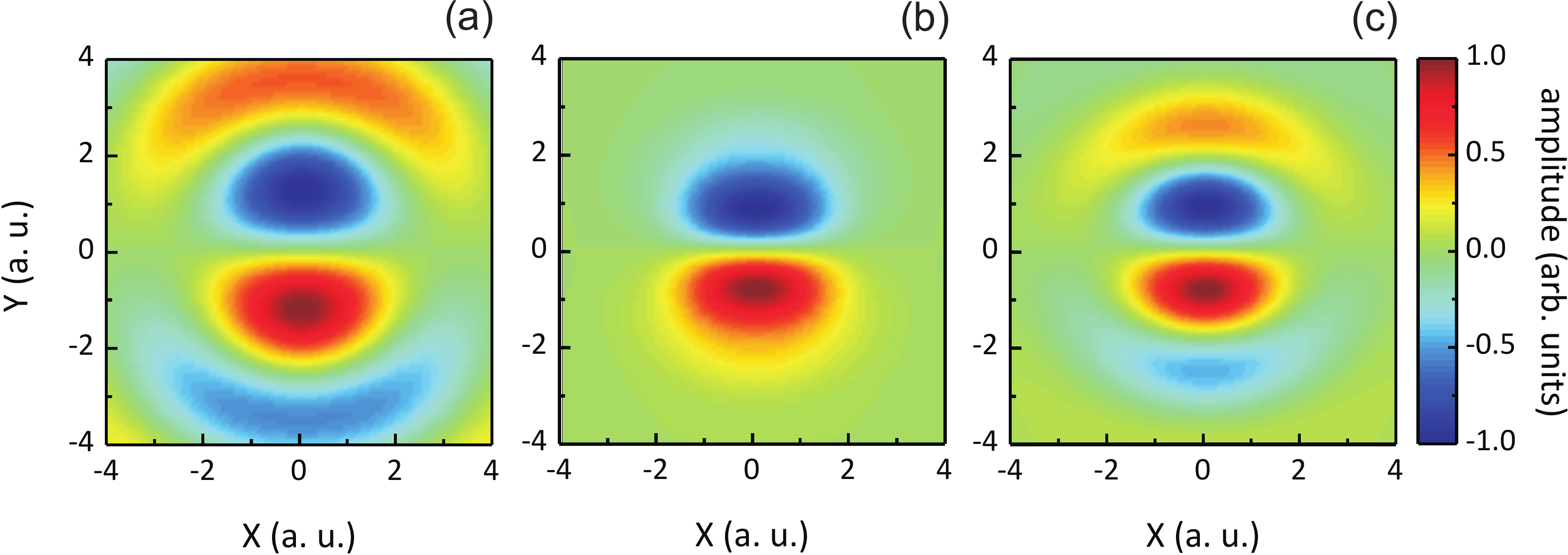}
\caption{a) Highest occupied molecular orbital of C$_{2}$H$_{2}$ as retrieved 
from the single molecule XUV emission map. (b) Highest occupied molecular 
orbital of C$_{2}$H$_{2}$ calculated with a quantum chemistry program 
\cite{dalton}. (c) C$_{2}$H$_{2}$ HOMO calculated considering the filtering in 
spectral domain corresponding to the experimental conditions. \label{homo_C2H2}}
\end{figure}

Differently from the case of N$_{2}$O, in C$_{2}$H$_{2}$ it is not possible to 
easily rule out the multi-electron contributions.
Because of the smaller cutoff energy, the experimental approach applied in the 
case of N$_{2}$O for the exclusion of multi-electron contribution is not feasible. 
Nevertheless the application of tomographic approach to the single molecule 
emission maps shown in figure \ref{SMXEM_C2H2} provides interesting results. We 
show in figure \ref{homo_C2H2}(a) the retrieved C$_{2}$H$_{2}$ HOMO. Also in 
this case, a comparison with the result calculated with a quantum chemistry 
program (see figure \ref{homo_C2H2}(b)) shows a good agreement in the overall 
shape of the orbital. Again the additional lobes are related to the limited 
harmonic range detected in the experimental acquisition, as can be seen in 
figure \ref{homo_C2H2}(c) where the orbital is calculated taking into account 
the spectral filtering.

\section{Conclusions\label{conclusions}}
Since the pioneering work of Itatani et al. on molecular orbital imaging, the 
impressive advances in laser technologies 
gave the access to new mid-IR sources for driving HHG and pushing the harmonic 
emission far towards the soft-X ray range.
These sources allowed the application of HHG spectroscopy to fragile molecules 
as hydrocarbons, which play as prototypes
for the study of ubiquitous phenomena in chemistry and material science.
In this work we showed the application of molecular orbital reconstruction based 
on HHG to non-trivial samples, such as N$_{2}$O and 
C$_{2}$H$_{2}$. These results, though requiring further improvements, 
demonstrate the capability of molecular orbital tomography and
represent the first step towards the imaging of dynamical processes in complex 
molecules.

\section*{Acknowledgements}
The research leading to these results has received funding from LASERLAB-EUROPE 
(grant agreement n° 284464, EC Seventh Framework Programme),
from ERC Starting Research Grant UDYNI (grant agreement n° 307964, EC Seventh 
Framework Programme)
and from the Italian Ministry of Research and Education (ELI project - ESFRI 
Roadmap).

\noindent\small
\bibliography{rsc}
\bibliographystyle{plain}

\end{document}